# Ralph: A Visible/Infrared Imager for the New Horizons Pluto/Kuiper Belt Mission

Dennis C. Reuter[1], S. Alan Stern[2], John Scherrer[3], Donald E. Jennings[1], James Baer[4], John Hanley[3], Lisa Hardaway[4], Allen Lunsford[1], Stuart McMuldroch[5], Jeffrey Moore[6], Cathy Olkin[7], Robert Parizek[4], Harold Reitsma[4], Derek Sabatke[4], John Spencer[7], John Stone[3], Henry Throop[7], Jeffrey Van Cleve[4], Gerald E. Weigle[3] and Leslie A.Young[7]

[1]NASA/GSFC, Code 693, Greenbelt, MD 20771
[2]Space Sciences and Engineering Division, Southwest Research Institute (SwRI), 1050 Walnut St., Suite 400, Boulder CO, 80302
[3]SwRI, 6220 Culebra Rd., San Antonio TX, 78228
[4]Ball Aerospace and Technology Corporation (BATC), 1600 Commerce St, Boulder, CO 80301
[5]SSG Precision Optronics, 65 Jonspin Rd., Wilmington MA, 01887
[6]NASA/Ames Research Center, MS 245-3, Moffett Field, CA 94035-1000
[7]Dept. of Space Studies, Southwest Research Institute (SwRI), 1050 Walnut St., Suite 400, Boulder CO, 80302

## ABSTRACT

The New Horizons instrument named Ralph is a visible/near infrared multi-spectral imager and a short wavelength infrared spectral imager. It is one of the core instruments on New Horizons, NASA's first mission to the Pluto/Charon system and the Kuiper Belt. Ralph combines panchromatic and color imaging capabilities with IR imaging spectroscopy. Its primary purpose is to map the surface geology and composition of these objects, but it will also be used for atmospheric studies and to map the surface temperature. It is a compact, low-mass (10.5 kg) power efficient (7.1 W peak), and robust instrument with good sensitivity and excellent imaging characteristics. Other than a door opened once in flight, it has no moving parts. These characteristics and its high degree of redundancy make Ralph ideally suited to this long-duration flyby reconnaissance mission.

## 1. INTRODUCTION

New Horizons, a flyby mission to the Pluto/Charon system and the Kuiper Belt, is the first in NASA's New Frontiers line of moderate-scale planetary missions and is the first mission to explore Pluto and its moons Charon, Nix and Hydra. Launched on January 19, 2006, it is scheduled for a closest approach of about 10,000 km on July 14, 2015. The scientific rationale for New Horizons and the overall mission planning are described in detail in several papers in this issue (Stern, Young *et al* and Fountain *et al*.). The New Horizons mission is led by Principal Investigator Alan Stern of the Southwest Research Institute of Boulder, CO and is managed by SwRI and the Johns Hopkins Applied Physics Laboratory in Laurel, MD. A core remote-sensing instrument on New Horizons is Ralph, a visible/NIR camera and infrared spectral mapper. The instrument's primary purpose is to measure surface characteristics, including geological processes, geomorphology, photometric properties, and surface composition. In addition, surface temperature will be inferred from the shapes and positions of well-established, thermally diagnostic reflectance spectral features in $H_2O$, $CH_4$, and $N_2$ ices. Ralph will also be used to measure haze optical depths (if present) and to search for rings and small satellites. This paper describes Ralph and specifies its characteristics (See also Reuter *et al*., 2005).

The Ralph instrument is mounted to the exterior of the New Horizons spacecraft (Fountain *et al*., this issue). Ralph consists of a single telescope that feeds two sets of focal planes: 1) the Multi-spectral Visible Imaging Camera (MVIC), a visible, near-IR imager and 2) the Linear Etalon Imaging Spectral Array (LEISA), a short-wavelength, IR, spectral imager. The telescope uses an unobscured, off-axis, three-mirror anastigmat design. The entire telescope assembly, including the three diamond turned mirrors, is constructed from grain aligned 6061-T6 aluminum. The optical mounts, housing and baffles are diamond turned from a single Al block. This combination of an all Al structure and optics is lightweight, athermal and thermally conductive. It ensures that the optical performance of the system is minimally sensitive to temperature and that thermal gradients are minimized. The highly baffled 75 mm aperture, VIS/IR telescope provides ample sensitivity at Pluto/Charon, while minimizing size and mass. The f/8.7 system's approximately 658 mm Effective Focal Length offers a good compromise between photometric throughput and alignment stability. Stray light control is improved by using a field baffle at an intermediate focus between the secondary and tertiary mirrors, and by using a Lyot stop at the exit pupil after the tertiary mirror. A dichroic beamsplitter transmits IR wavelengths longer than 1.1 µm to LEISA and reflects shorter wavelengths to MVIC.

MVIC is composed of 7 independent CCD arrays on a single substrate. It uses two of its large format (5024x32 pixel) CCD arrays, operated in time delay integration (TDI) mode, to provide panchromatic (400 to 975 nm) images. Four additional 5024x32 CCDs, combined with the appropriate filters and also operated in TDI mode, provide the capability of mapping in blue (400-550 nm), red (540-700 nm), near IR (780 – 975 nm) and narrow band methane (860 – 910 nm) channels. TDI operates by synchronizing the parallel transfer rate of each of the CCDs thirty-two 5024 pixel wide rows to the relative motion of the image across the detector's surface. In this way, very large format images are obtained as the spacecraft scans the FOV rapidly across the surface. The presence of 32 rows effectively increases the integration time by that same factor, allowing high signal-to-noise measurements. The FOV of a single MVIC pixel is 20x20 µradian$^2$. The panchromatic (pan) channels of MVIC will be used to produce hemispheric maps of Pluto and Charon at a double sampled spatial resolution of 1 km$^2$ or better. The static FOV of each of the TDI arrays is 5.7°x0.037°. In addition to the TDI arrays, MVIC has a 5024x128 element, frame transfer panchromatic array operated in staring mode, with an FOV of 5.7°x0.15°. The primary purpose of the framing array is to provide data for optical navigation (OpNav) of the spacecraft.

LEISA is a wedged filter infra-red spectral imager that creates spectral maps in the compositionally important 1.25-2.5 micron short wave infrared (SWIR) spectral region. It images a scene through a wedged filter (linear variable filter, Rosenberg *et al*., 1994) placed about 100 µm above a 256x256 pixel Mercury Cadmium Telluride (HgCdTe) detector array (a PICNIC array). An image is formed on both the wedged filter and the array simultaneously (there is less than 5% spectral broadening by the f/8.7 beam). LEISA forms a spectral map by scanning the FOV across the surface in a push broom fashion, similar to that of the MVIC TDI channels. The frame rate is synchronized to the rate of the scan, so that a frame is read out each time the image moves by the single pixel IFOV. The LVF is fabricated such that the wavelength varies along one dimension, the scan direction. The difference between a LEISA scan and a TDI scan is that in LEISA the row-to-row image motion builds up a spectrum while in TDI the motion increases the signal over a single spectral interval. The filter is made in two segments. The first covers from 1.25 to 2.5 microns at an average spectral resolving power ($\lambda/\Delta\lambda$) of 240. This section of LEISA will be used to obtain composition maps. The second segment covers 2.1 to 2.25 microns with an average spectral resolving power of 560. It will be used to obtain both compositional information and surface temperature maps by measuring the spectral shape of solid $N_2$.

The MVIC and LEISA components of Ralph were originally developed in 1993 for what was then called the "Pluto Fast Flyby" mission using grants from NASA's "Advanced Technology Insertion" (ATI) project. At the time, they were combined with a UV mapping spectrometer (also developed under the ATI grant) into a fully integrated remote-sensing package called HIPPS (Highly Integrated Pluto Payload System, Stern *et al*, 1995). For the New Horizons mission, HIPPS evolved into the Pluto Exploration Remote Sensing Instrument (PERSI), in which the UV spectrometer, now named Alice, was decoupled from the MVIC and LEISA components. This allowed the UV and Vis/IR optics to be separately optimized and reduced the chances of contamination of the sensitive UV optics. Versions of Alice are flying on the Rosetta comet orbiter mission and have been chosen for the Lunar Reconnaissance Orbiter (see Stern *et al*., in this issue and references therein). Instruments based on the ATI LEISA concept have flown on the Lewis mission (Reuter *et al*., 1996) and the EO-1 mission (Reuter *et al*., 2001, Unger *et al*, 2003). The Lewis spacecraft prematurely re-entered the atmosphere before any instrument aboard could take data, but the version of LEISA aboard EO-1 provided numerous images. Now that the UV Alice spectrometer is a separate entity from the MVIC/LEISA sub-assembly, the latter is named Ralph in honor of Ralph and Alice Kramden of "Honeymooner's" fame. New Horizons is the first mission on which the Ralph instrument has flown.

Ralph is a joint effort of the Southwest Research Institute (SwRI, San Antonio, TX and Boulder, CO which is the home institution of Alan Stern, the Ralph Principal Investigator), Ball Aerospace and Technologies Corp. (BATC, Boulder, CO) and NASA's Goddard Space Flight Center (GSFC, Greenbelt, MD).

## 2. RALPH SCIENCE OVERVIEW

The scientific rationale for the New Horizons mission to the Pluto system (and beyond into the deeper Kuiper Belt) is given in detail in another paper in this volume (Young *et al.*) and will not be repeated at length here. In brief, the Kuiper Belt is an extended disk containing numerous primordial bodies whose planetary evolution was arrested early in Solar System formation. In essence, Kuiper Belt Objects (KBOs) are the "fossils" of planetary evolution and the Kuiper Belt is the prime "archeological site" in the Solar System. Pluto is among the largest known KBOs and is a full-fledged dwarf planet. Because Charon is approximately half of Pluto's size, the center of mass of the Pluto-Charon system lies between the two objects, making it a true binary system. The first exploration of the Pluto-Charon system and the Kuiper Belt is both scientifically and publicly exciting. It will provide invaluable insights into the origin of the outer solar system and the ancient outer solar nebula. It will explore the origin and evolution of planet–satellite systems and the comparative geology, geochemistry, tidal evolution, atmospheres, and volatile transport mechanics of icy worlds.

The Ralph instrument will play a leading role in this exploration. It directly addresses two of the three Group 1, or primary, mission requirements (see Stern, this volume): 1) to characterize the global geology and morphology of Pluto and Charon and 2) to map the surface composition of Pluto and Charon. It also contributes to the third Group 1 requirement of searching for atmospheric haze. High spatial resolution (≤1 km/line pair) panchromatic maps generated by the MVIC component of Ralph will be used to address the first Group 1 requirement. These maps will be obtained for the hemisphere visible at closest approach, and will address many of the outstanding questions about these bodies. What is their cratering history? What types of structures are found on their surfaces? What is the spatial variability and scale size of surface features? What is the effect of seasonal volatile transport on the "smoothness" of surface features? Answers to these questions will revolutionize our understanding of the formation and evolution of the Pluto/Charon system.

The second Group 1 goal will be addressed both by MVIC and by LEISA. LEISA will obtain hemispheric maps in the 1.25 to 2.5 µm spectral region with an average resolving power ($\lambda/\Delta\lambda$) of 240 and a spatial resolution of less than 10 km. Similarly, MVIC will provide hemispheric surface color maps and maps of surface methane ($CH_4$) at even higher spatial resolutions. Pluto's surface is known to contain the species $CH_4$, $N_2$, and CO, while Charon's is primarily $H_2O$ but is also likely to contain ammonium hydrates. LEISA's hemispheric maps will allow us to address questions pertaining to composition. What is the surface distribution of the main species? Are there areas of pure frost and mixed areas? What is the effect of seasonal transport? Are there more complex species in selected regions of the surface? Is there a connection between geology and composition? Answers to these questions will significantly advance our understanding of the chemical and physical processes that occur on icy objects and of the processes that occurred in the cold outer regions of our solar system during its formation. Figure 1 shows simulated spectra for Pluto and Charon. As is evident from this figure, there is a wealth of information to be gleaned from this spectral region even from globally averaged spectra. LEISA's spectral maps will permit the correlation of composition with both geology and atmospheric transport of volatiles. They will also enable the study of Pluto's crustal composition where craters or other windows into the interior so permit.
.
In addition to the primary, Group 1 objectives, Ralph will address numerous Group 2 and Group 3 measurement goals. These include: obtaining stereo images of Pluto and Charon (MVIC), mapping the terminators (MVIC), obtaining high resolution maps in selected regions (MVIC and LEISA), refining the bulk parameters and orbits of the Pluto system and searching for rings and additional satellites (MVIC). LEISA will use its high resolution ($<\lambda/\Delta\lambda> = 560$) 2.1 to 2.25 µm segment to obtain surface temperature maps of Pluto employing a technique that relates temperature to the spectral shape of the $N_2$ transition near 2.15 µm. This technique is particularly sensitive for temperatures near 35 K, the temperature at which $N_2$ undergoes a transition from α phase to β phase (Grundy *et al.*, 1993). Thirty five K is close to the predicted surface temperature of Pluto at the time of the flyby in 2015. For Pluto, additional temperature information will be obtained from the band shape of the $CH_4$ features measured using the lower resolution LEISA filter. For Charon which does not have a prominent $N_2$ band, reasonably accurate temperature maps can be deduced from the shape of the water bands observed with the lower resolution LEISA filter. These secondary and tertiary objectives, while not mission critical like the Group 1 goals, will add substantially to our understanding of Pluto and Charon. MVIC panchromatic and color maps and LEISA spectral maps will also be obtained of Nix and Hydra, two recently discovered satellites in the Pluto system.

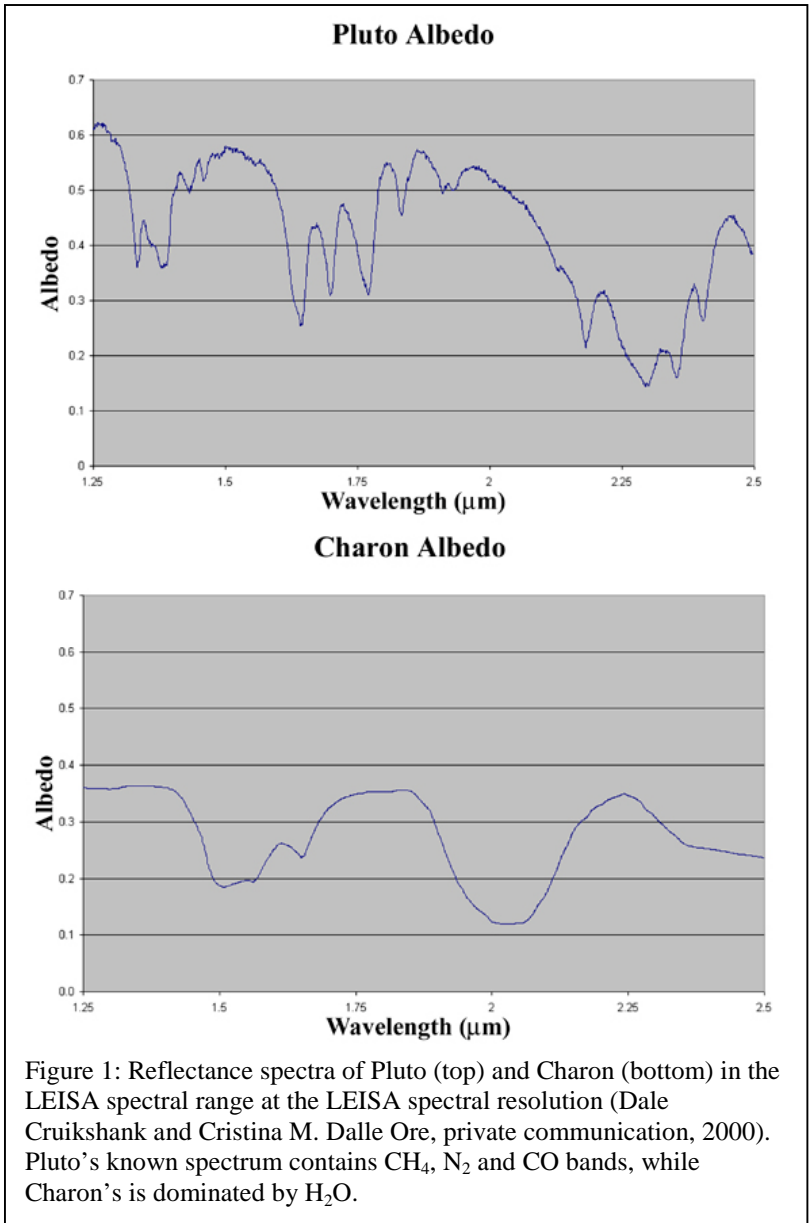

Figure 1: Reflectance spectra of Pluto (top) and Charon (bottom) in the LEISA spectral range at the LEISA spectral resolution (Dale Cruikshank and Cristina M. Dalle Ore, private communication, 2000). Pluto's known spectrum contains $CH_4$, $N_2$ and CO bands, while Charon's is dominated by $H_2O$.

Table 1 summarizes the science objectives that determined the Ralph design, the measurement strategies that address these objectives and the derived instrument performance requirements. Except for the spectral resolution and coverage of the high-resolution segment of LEISA, the performance requirements were determined by the need to address the Group 1 goals. The high-resolution LEISA segment was added specifically to address the Group 2 goal of mapping Pluto's surface temperature, but it will prove useful in the surface composition mapping as well. The MVIC framing camera will be used to perform optical navigation. This gives rise to the additional requirement that it be capable of measuring a $10^{th}$ magnitude star with a signal-to-noise ratio of 7 in a 0.25 second exposure.

## 3. OPTO-MECHANICAL DESIGN

Figure 2 shows a model of Ralph and a picture of the assembled instrument before addition of the final MLI. The major elements are labeled in the model. The mass of the instrument is 10.5 Kg and the maximum peak power load is 7.1 Watts. The low power and mass are especially important for the New Horizons mission where both of these resources are at a premium. As shown in Fig. 2, Ralph has two assemblies, the telescope detector assembly (TDA) and the electronics assembly. The TDA consists of the telescope optical elements, the baffling, the MVIC and LEISA focal planes, the two-stage passive radiator that cools the focal planes and the flat fielding Solar Illumination Assembly (SIA). The aperture of the TDA was closed by a one-time use door with a partially transmitting window (about 20% throughput). The door protected the optics from contamination prior to and during launch and protected the focal planes from accidental solar exposure during the early flight stage. The door was opened when the spacecraft was 2.3 AU from the sun and can not be closed. The TDA is mounted to the spacecraft by thermally isolating titanium flexures. The in-flight temperature of the TDA is about 220 K, The temperature of the electronics box, which is mounted directly to the spacecraft, is about 290 K. The low temperature of the TDA reduces the conductive and radiative thermal load on the focal planes. It also limits the background signal at the long wavelength end of LEISA. The inner stage of the externally mounted passive radiator cools the LEISA detector to < 130 K. The outer annulus maintains the MVIC CCDs at temperatures below 175 K and lowers the temperature of an f/2.4 cold shield for LEISA to below 190 K. The 75 mm aperture, 657.5 mm focal length, f/8.7 telescope provides good image quality over the 5.7°x1.0° field of view spanned by the MVIC and LEISA arrays. The instrument parameters for Ralph are summarized in Table 2. Figure 3 shows a model of the TDA interior with a raytrace diagram.

**Table 1: Science Objectives and Derived Instrument Requirements**

| Science Objective | Measurement Strategy | Derived Instrument Requirements | | | |
|---|---|---|---|---|---|
| | | Spectral coverage | Resolution | Image Quality | Signal-to-noise |
| Global geology and morphology of Pluto/Charon | Panchromatic images: spatial resolution of 1 km/line pair | 400 – 975 nm | N/A | MTF ≥0.15 @ 20 cycles/milliradian | 50 (33 AU, 0.35 I/F) |
| Map the surface composition of Pluto/Charon | Color images: spatial resolution <10 km/line pair | 400 – 550 nm (blue) 540 – 700 nm (red) 780 – 975 nm (NIR) 860 – 910 nm ($CH_4$) | N/A | No additional requirement. | 50 (33 AU, 0.35 I/F) 50 (33 AU, 0.35 I/F) 50 (33 AU, 0.35 I/F) 15 (33 AU, 0.35 I/F) |
| Map the surface composition of Pluto/Charon | SWIR spectral images: spatial resolution <10 km | 1.25 – 2.5 µm | $\lambda/\Delta\lambda \geq 250$ | No additional requirement. | 32 (1.25 µm; Pluto) 27 (2.00 µm; Pluto) 18 (2.15 µm; Pluto) |
| Map Pluto's surface temperature | High spectral resolution images in the 2.15 µm $N_2$ band | 2.10 – 2.25 µm | $\lambda/\Delta\lambda \geq 550$ | No additional requirement. | No additional requirement. |

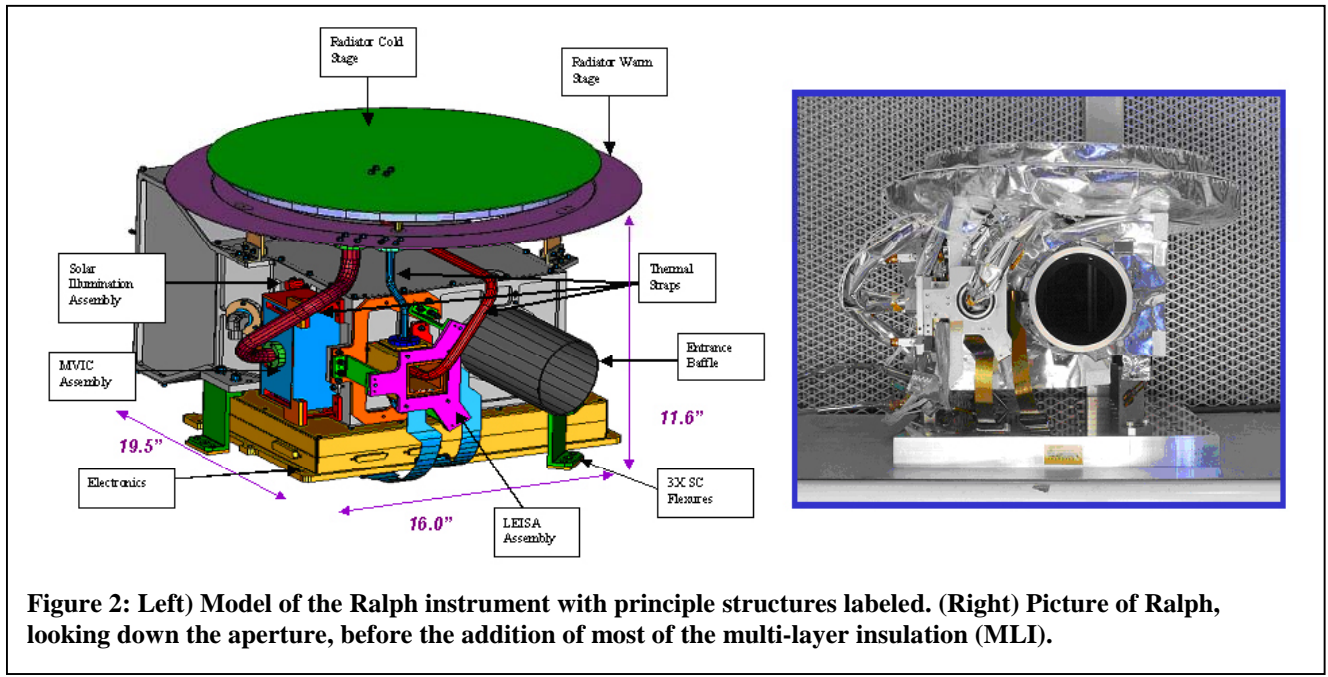

**Figure 2: Left) Model of the Ralph instrument with principle structures labeled. (Right) Picture of Ralph, looking down the aperture, before the addition of most of the multi-layer insulation (MLI).**

**3.1 The MVIC Focal Plane**

The MVIC focal plane assembly consists of a customized CCD array provided by E2V Corp. of Chelmsford, England. It is mounted to a heat sink plate and placed directly behind a "butcher block" filter assembly. The array has six identical 5024x32 pixel (5000x32 pixel photoactive area) TDI CCDs and one 5024x264 pixel (5000x128 pixel photoactive area) frame transfer CCD on a single substrate. Figure 4 shows a schematic of the array indicating the positions of the CCDs and a picture of the actual flight array. The long dimension of 5000 elements was chosen because to obtain a 1 km/line pair image across Pluto's 2300 km diameter in a single scan requires at least 4600 pixels. The remaining 400 pixels allow for pointing inaccuracies and drift during the scan. All MVIC pixels are 13x13µm$^2$.

The frame transfer array consists of two regions; the 5024x128 pixel image gathering area, and a 5024x136 pixel image storage area. The extra eight rows in the image storage area contain injection charge that reduces charge traps. For both the TDI and framing arrays, the extra 24 dark pixels (12 on each side of the 5000 pixel active region) are used as reference pixels and for injected charge. A front-side illuminated CCD is used to optimize the imaging qualities of the array. The filter, which was provided by Barr Associates, is mounted about 700 microns above the surface of the array. It has five segments, four with the passbands described in Table 2 placed directly over the four CCDs forming the color segment of MVIC. The remaining two TDI CCDs and the frame transfer array are overlain by a clear filter so that the focus position is the same for all seven arrays. In TDI mode, the spacecraft is rotated to scan the image of each segment of the surface across the focal plane in a pushbroom fashion. The entire surface may be imaged in this way. The nominal rotation rates are about 1600 µrad/sec for the pan band and about 1000 µrad/sec for the color bands. These correspond to integration times of about 0.4 and 0.6 seconds respectively. The clocking of each 5024 element pixel row is synchronized with the spacecraft motion using attitude control knowledge obtained from the spacecraft so that the effective integration time is 32 times the row transfer period. In this way, the signal-to-noise ratio of the observations is increased while the time required to obtain a full image is reduced. TDI mode takes advantage of the ability of the spacecraft to scan smoothly in attitude and does not need the multiple pointing operations that a mosaic of framing images would require. In flight, the CCD clocking rate errors in the TDI (along scan) direction have been shown to cause less than ¼ of a pixel of excess image "smear" for integration times of 0.7 seconds or less.

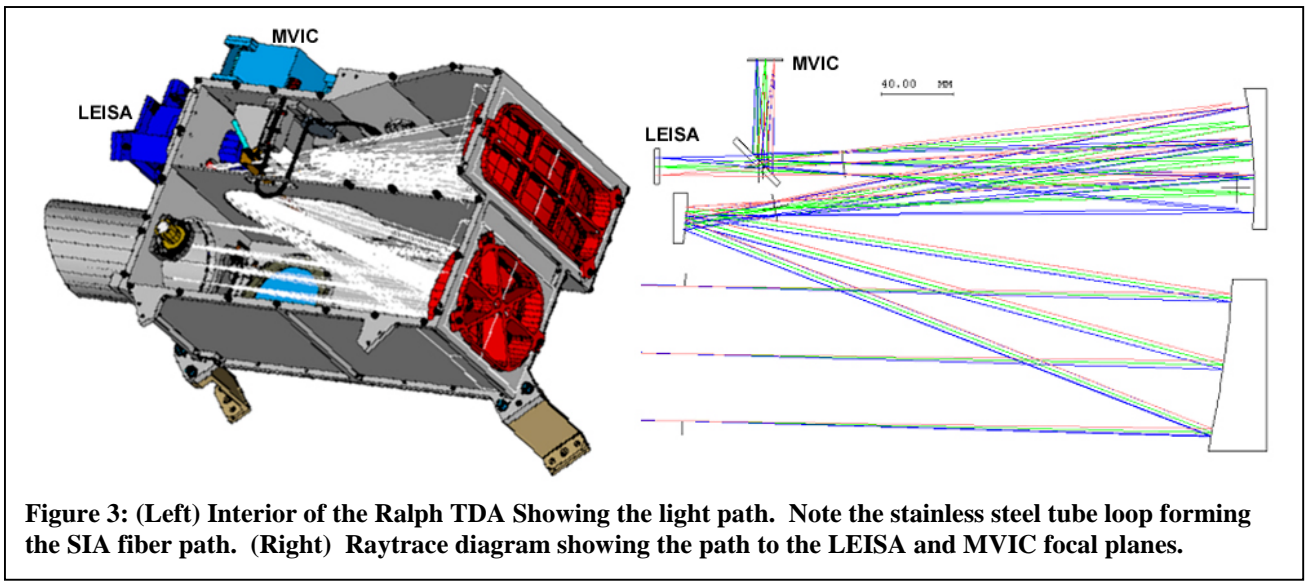

**Figure 3:** (Left) Interior of the Ralph TDA Showing the light path. Note the stainless steel tube loop forming the SIA fiber path. (Right) Raytrace diagram showing the path to the LEISA and MVIC focal planes.

### 3.2 The LEISA Focal Plane

The LEISA IR spectral imager also works in pushbroom mode, except in this case, the image motion is used to scan a surface element over all spectral channels. The wedged filter effectively makes each column of the array

**Table 2: Ralph Instrument Parameters**

Mass: 10.5 Kg
Power: 7.1 Watt (maximum)
Telescope Aperture: 75 mm
Focal Length: 657.5 mm
f/#: 8.7
**MVIC**: Time Delay and Integrate (TDI) and framing arrays
  2 Redundant 5024x32 Pixel Panchromatic TDI CCDs (400 – 975 nm)
  Four 5024x32 Pixel Color TDI CCDs
    Blue (400 – 550 nm)
    Red (540 – 700 nm)
    NIR (780 – 975 nm)
    Methane (860 – 910 nm)
  5024x128 Frame Transfer Pan CCD
  13µm x 13µm pixels
  Single pixel Field of View: 19.77µrad x 19.77µrad
  TDI array FOV: 5.7°x0.037°
  Framing camera FOV: 5.7°x0.146°
  Focal plane temperature: <175 K
  Pan TDI rate: 4 – 84 Hz
  Color TDI rate: 4 – 54 Hz
  Frame transfer integration time: 0.25 – 10 sec.
**LEISA**: 256x256 element HgCdTe array operated in pushbroom mode.
  40µm x40µm pixels
  Single pixel Field of View: 60.83µrad x 60.83µrad
  FOV: 0.9°x0.9°
  Focal plane temperature: <130 K
  Filter segment 1 (1.25 – 2.5 µm) average resolving power (λ/Δλ): 240
  Filter segment 2 (2.1 – 2.25 µm) average resolving power (λ/Δλ): 560
  Frame rate: 0.25 – 8 Hz

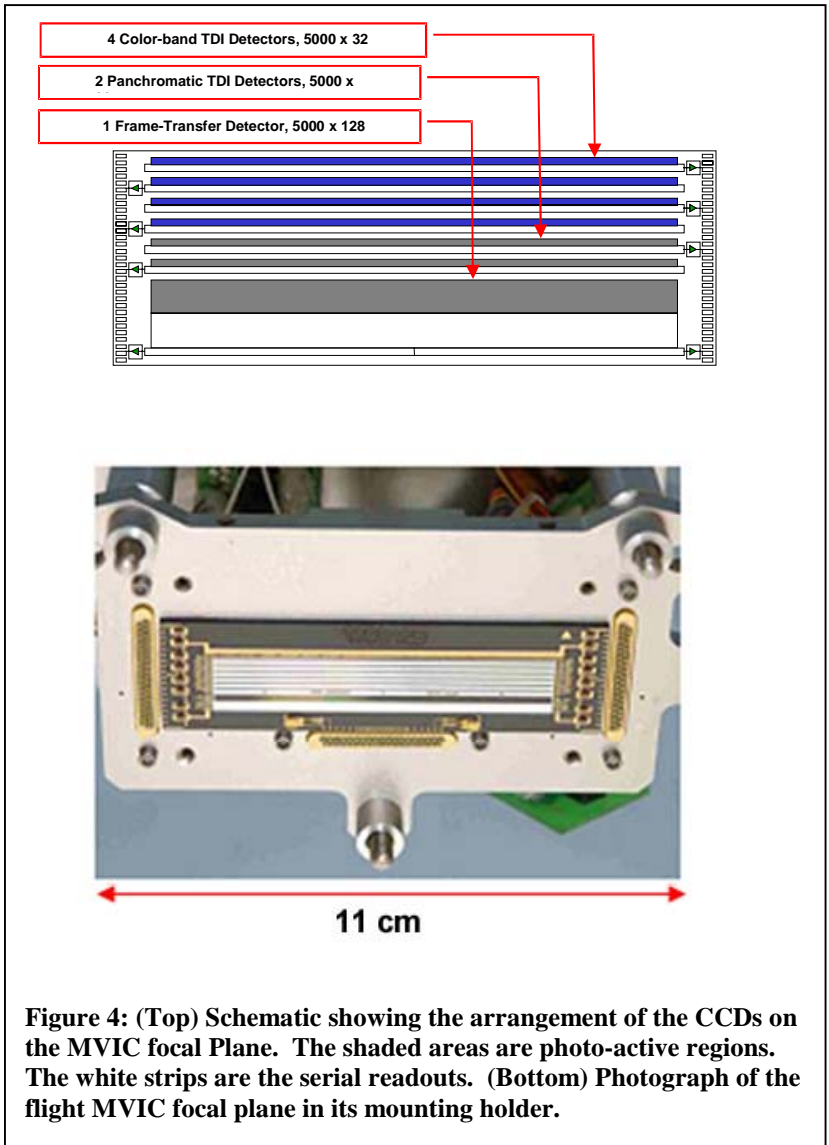

**Figure 4:** (Top) Schematic showing the arrangement of the CCDs on the MVIC focal Plane. The shaded areas are photo-active regions. The white strips are the serial readouts. (Bottom) Photograph of the flight MVIC focal plane in its mounting holder.

responsive to only a narrow band of wavelengths, so that, conceptually, the filter may be considered as consisting of 256 adjacent narrow band filters. As with MVIC, the image is scanned over the LEISA focal plane by rotating the spacecraft. The nominal rotation rate is about 120 μrad/sec for a frame rate of 2 Hz. Here again, the frame rate is synchronized to the spacecraft-measured rotation rate, so that the image moves one column per frame.

The LEISA detector is a 1.2 to 2.5 μm HgCdTe PICNIC array, supplied by Rockwell Scientific Corporation of Camarillo CA. The array is a 256x256 pixel array and each pixel is $40x40 \mu m^2$ in area. However, several modifications were made to the standard PICNIC array. The HgCdTe was grown on a CdTe substrate using Molecular Beam Epitaxy (MBE) to provide good lattice matching and low dark currents. The detector was bump bonded to a standard PICNIC multiplexer and the resulting hybrid was mounted to molybdenum pad. This process reduces mechanical stress induced during cooling to operational temperature. It is estimated that the assembly can safely undergo at least 1000 thermal cycles. The electrical interface to the array is provided by two ribbon cables and a multilayer fan-out board that were fabricated into a single element. The LEISA array is back illuminated, but the substrate has been thinned from 800 μm to 200 μm so that the active area of the array is significantly closer to the surface than is usual. This puts both the filter and the array within the depth of focus. The filter, supplied by JDSU Uniphase/ Optical Coating Laboratories Inc. of Santa Rosa CA, was made in two segments. The first, covering from 1.25 to 2.5 μm at a constant resolving power (constant $\Delta\lambda/\lambda$) of about 240, provides information primarily for surface composition mapping. The second, covering from 2.1 to 2.25 μm at a constant resolving power of about 560, uses temperature dependent changes in the spectral structure of solid $N_2$ near the α to β phase transition at 35 K to provide surface temperature maps. In both segments, a constant resolving power is achieved by making the transmitted wavelength depend logarithmically on position. The two segments were bonded together to form a single filter element. This filter was, in turn, bonded into a holder and mounted such that the filter surface is about 100 μm above the surface of the array. The refractive index of the array is approximately 2.7 so that the total optical path between the filter and photo-active area of the array is less than 200 μm. In this distance, the f/8.7 beam spreads about 0.5 pixel, so when the focus position is optimized between the array and filter surface, the convolved image smear is about 0.04 pixel. A picture of the array and the complete focal plane assembly is shown in Figure 5.

**3.3 The Solar Illumination Assembly (SIA)**

The SIA is a second input port whose FOV is along the spacecraft antenna pointing direction at 90 degrees with respect to the main aperture. It is designed to provide diffuse solar illumination to both the MVIC and LEISA focal planes. In practice, it illuminates all of the LEISA array and about 3000 pixels of the each MVIC array with a reproducible pattern that can be used for determining the stability of the pixel-to-pixel response (flat-fielding) during the mission. The SIA consists of a small fused silica lens (4 mm aperture, 10 mm focal length) that images the sun onto the input end of a 125 μm core fiber. The output end of the fiber illuminates a pair of lenses, which are directly under the Lyot stop (the exit pupil) and about 10 cm from both focal planes. To obtain an SIA measurement, the spacecraft is oriented so that the SIA lens images the sun onto the fiber. At Pluto, the diameter of the image of the Sun on the fiber would be about 50 μm. This is significantly larger than the diffraction limited image size because of chromatic aberration in the single element lens. Nevertheless, the solar image underfills the fiber, so the intensity level is relatively insensitive to pointing errors. The fiber is about 10 cm long and is contained in a stainless steel tube. It is more than 50% transmitting over the full spectral band from 0.4 to 2.5 μm. There is a second fiber in the SIA with an attenuation factor of about 40 that can be used for flat fielding nearer the Sun (e.g. at Jupiter).

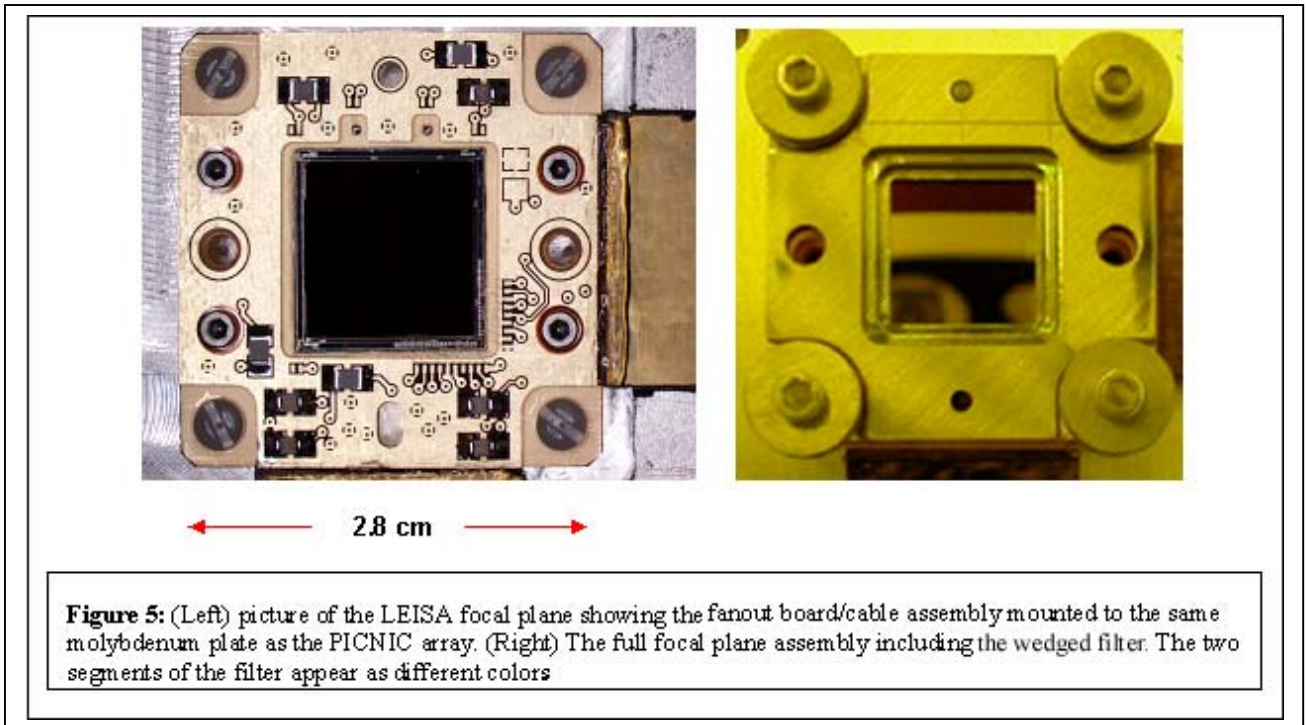

**Figure 5:** (Left) picture of the LEISA focal plane showing the fanout board/cable assembly mounted to the same molybdenum plate as the PICNIC array. (Right) The full focal plane assembly including the wedged filter. The two segments of the filter appear as different colors

A second possible use of the SIA is as a solar limb viewing port. In this mode, an atmospheric spectrum can be measured as a function of tangent height as a planet's atmosphere occludes the Sun. Vertical spectral profiles would be obtained using this capability. To increase sensitivity in sparse atmospheres, such as Pluto's, the spectra from all rows may be summed into a single spectrum. The SIA is co-aligned with the Alice solar occultation (SOC) port (Stern *et al.*, this volume).

### 4.0 ELECTRONICS

The Ralph control electronics consist of three boards; detector electronics (DE), command and data handling (C&DH) and a low voltage power supply (LVPS). These are contained within an electronics box (EB) mounted directly to the spacecraft, below the TDA (see Fig. 2), and operate essentially at the spacecraft surface temperature, which is near ambient. The DE board provides biases and clocks to both focal planes, amplifies the signals from the arrays and performs the A/D conversion of the imaging data. The science data are converted with 12 bits per pixel. The C&DH board interprets the commands, does the A/D conversion of the low speed engineering data and provides both the high speed imaging data interface and the low speed housekeeping data interface. The LVPS converts the 30V spacecraft power to the various voltages required by Ralph.

In a long duration mission such as New Horizons, reliability of the electronics is of paramount importance, particularly for a core instrument that addresses all three Group 1 objectives. To ensure that Ralph is robust, almost all of the electronics are redundant. As illustrated in Figure 6, Ralph can operate on two separate sides (side A or B) which have very few components in common. The only common elements are: 1) the relays that choose whether side A or side B is to be powered, 2) The arrays themselves and 3) the interface to the spacecraft. However, the spacecraft interface has two identical circuits and is inherently redundant. For MVIC, the potential single point failure mode of the array is mitigated by dividing the six TDI arrays into two groupings, each containing two color CCDs and one panchromatic CCD. The first grouping is composed of a pan band and the red and $CH_4$ channels. The second grouping is composed of the other pan band and the blue and NIR channels. If either group should fail, the other would still be able to meet the science requirement of observations in 2 color bands and 1 panchromatic band. LEISA has 4 outputs corresponding to the four 128x128 quadrants of the array. If any one quadrant should fail, all science can still be completed. The same is true for the four out of six possible two-quadrant failures that still have active pixels at all wavelengths.

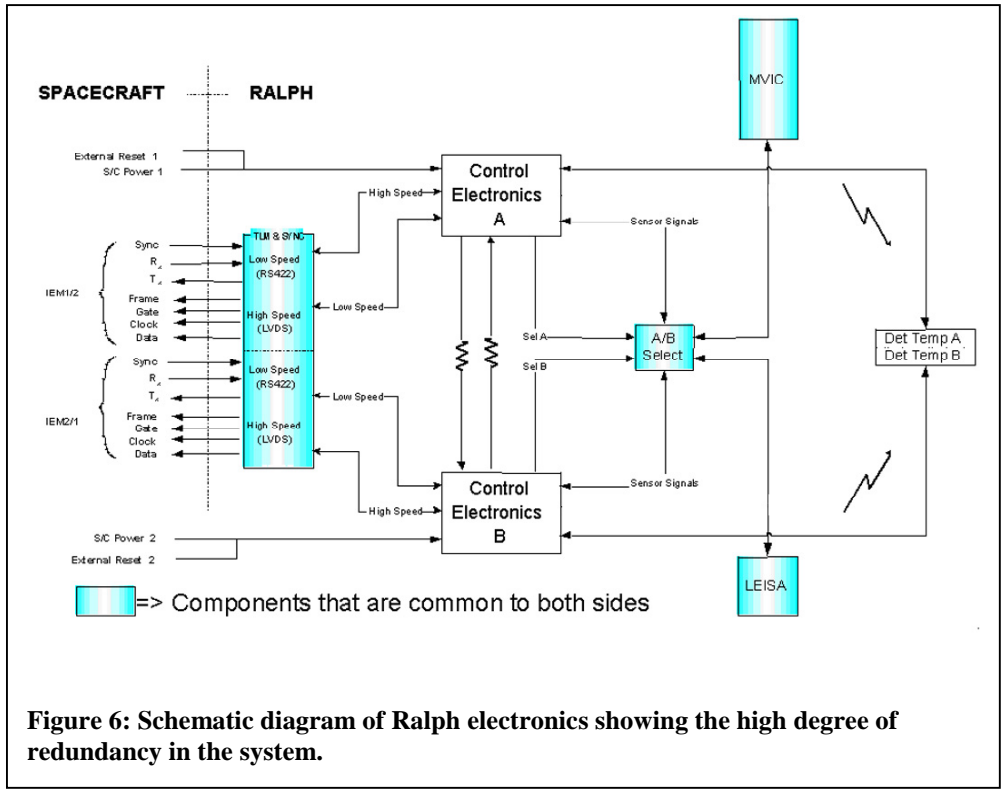

**Figure 6: Schematic diagram of Ralph electronics showing the high degree of redundancy in the system.**

MVIC always produces image data in correlated double sample (CDS) mode in which the reset level is subtracted from the integrated level. LEISA can send either CDS data, which is its standard operating mode, or "raw" data that contains both the reset levels and the integrated levels. The "raw" mode produces twice the data volume of the CDS mode and is used to set the LEISA A/D converter offset level for each quadrant of the array. The same offset is used in both CDS and "raw" modes. The selectable offset compensates for voltage drifts in the analog signal train over the life of the mission, maintaining dynamic range without sacrificing signal resolution (increasing quantization noise). For both focal planes the measured spacecraft rotation rate is fed back to the instrument to optimize the TDI or frame rate. That is, after a scan has been initiated the spacecraft determines the actual rotation rate and sends that information to Ralph, which uses it to calculate a TDI or frame rate such that the image moves a single row during a clock period. This reduces smear in the along-track direction.

For both MVIC and LEISA, the dominant noise source at low light levels is the system electronics noise, including array read noise. For MVIC this is about 30 e- (~ 200 $\mu V$) and for LEISA it is about 50 e- (~ 100 $\mu V$). The overall average gain for MVIC is about 58.6 e-/DN (Digital Number, or least significant bit), while for LEISA it is approximately 11 e-/DN.

## 5.0 PRE-LAUNCH INSTRUMENT CHARACTERIZATION

An extensive pre-launch program of performance verification measurements was carried out for Ralph at both the component level and the full instrument level. The component level characterization included measurements of the wavelength dependent quantum efficiency for the MVIC and LEISA array/filter assemblies and measurements of the wavelength dependence of the other optical elements (i.e. reflectance of the mirrors, transmission of the filters and throughput of the beamsplitter). Full instrument level testing was carried out under spaceflight-like conditions in a thermal vacuum chamber at BATC. The primary performance characteristics verified in these tests were relative system throughput (relative radiometric sensitivity) and image quality. The directional characteristics of the SIA were also measured.

**5.1 Component Level Measurements**

*LEISA Spectral Lineshape*: The instrument line shape was determined for the LEISA filter/array focal plane assembly over the entire 1.25 to 2.5 µm band by using a combination of multi-order grating and narrow band filter measurements. In this way a pixel-by-pixel table of the central wavelength, resolving power and out-of-band transmission was generated. Figure 7a shows an example of the readout along a single LEISA row when the focal plane assembly was illuminated using the output of a grating monochromator. The first five peaks correspond to orders 6 through 10 of the grating. The intensity decrease is primarily caused by the spectral shape of the source and by the decreasing efficiency of the grating at higher order. The line at 2.2 µm occurs in both segments of the filter. Figure 7b shows the intensity measured at a single pixel as the grating is scanned in small wavelength increments (0.33 nm). This instrument lineshape is representative of all pixels and is approximately gaussian. At this wavelength (2165 nm), the full width at half maximum (FWHM) is 8.3 nm, giving a resolving power of 260. Figure 8 shows the average resolving power for both filter segments of LEISA generated using the measurement technique described above. As may be seen from this figure, the average resolving power for the lower resolution segment is 240, with variations of 10 to 15 percent lower and higher. This is slightly below the requirement of 250. The resolution determines the reliability with which pure materials may be differentiated from mixtures and the accuracy of temperatures determined from line shapes. The slight decrease in accuracy in the lower resolution regions will not have significant scientific impact. Temperatures will be determined most accurately using the high-resolution segment. As required, the resolving power of the higher-resolution segment is greater than 550 in the region of the 2.15-μm $N_2$ band.

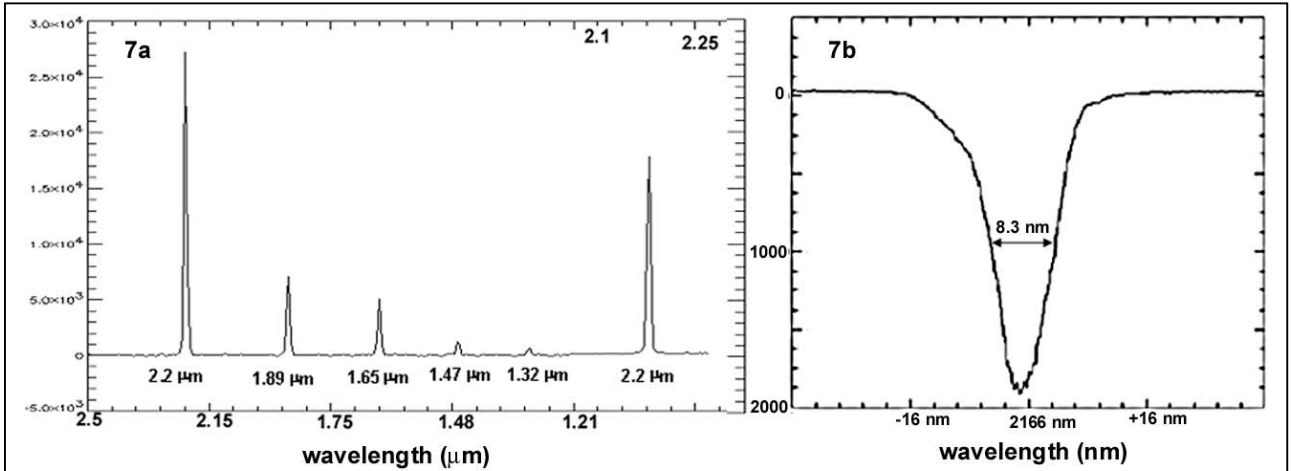

**Figure 7**: (a) A single row of LEISA showing multiple orders of a grating. Note the non-linear (logarithmic) wavelength scale and the presence of the 2.2 µm order in both the low resolution (2.5 to 1.25 µm, λ/Δλ ~ 240) and the high resolution (2.1 to 2.25 µm, λ/Δλ ~ 560) filter segments. (b) Wavelength dependence of transmitted intensity (instrument lineshape function) at a single pixel as the wavelength is varied from 2133 to 2200 nm in 0.3 nm steps.

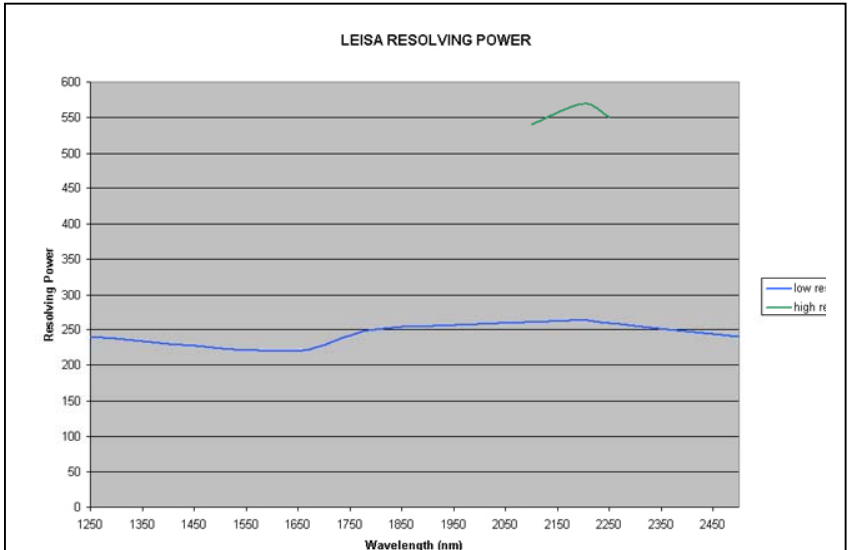

**Figure 8:** Measured resolving power (λ/Δλ) of the LEISA array/filter assembly for the lower-resolution (1.25 – 2.5 µm) and higher-resolution (2.1 – 2.25 µm) segments.

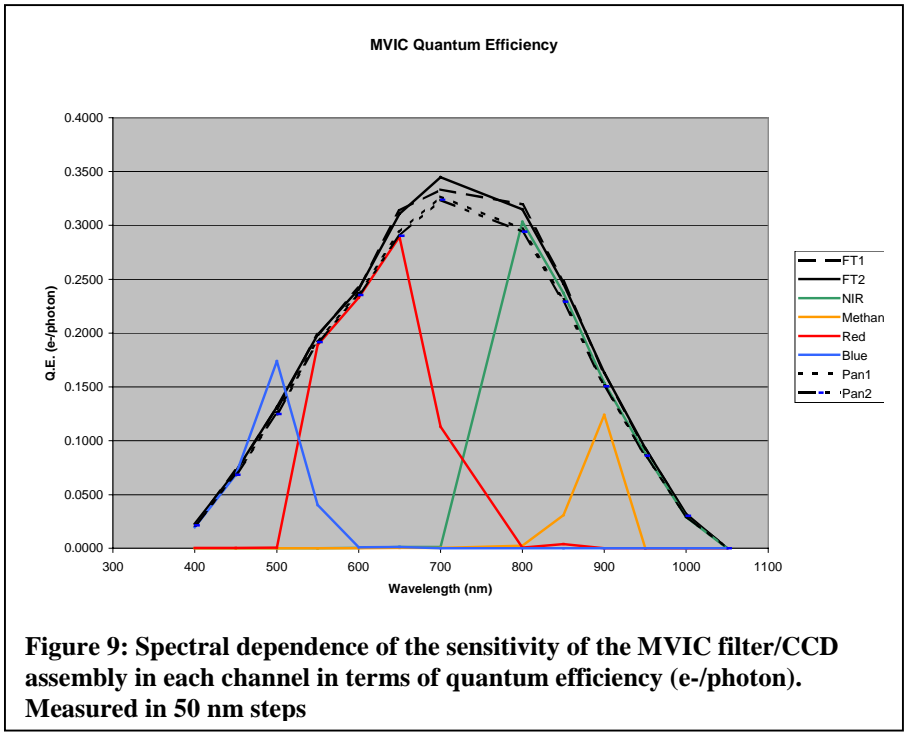

**Figure 9: Spectral dependence of the sensitivity of the MVIC filter/CCD assembly in each channel in terms of quantum efficiency (e-/photon). Measured in 50 nm steps**

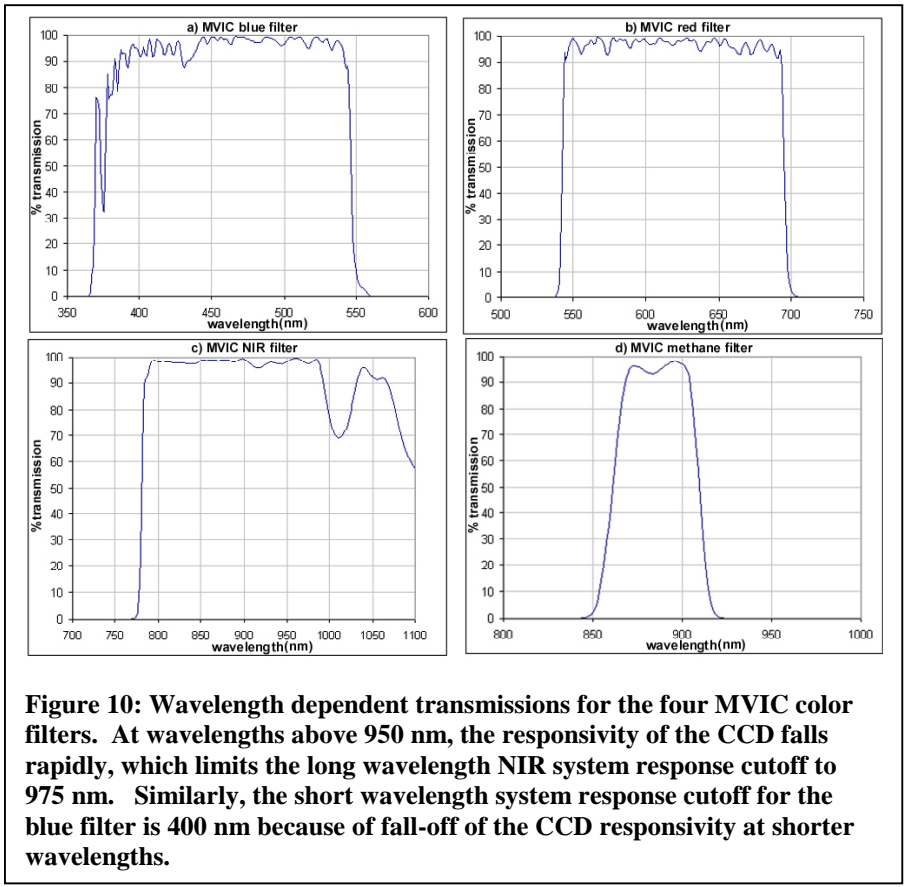

**Figure 10: Wavelength dependent transmissions for the four MVIC color filters. At wavelengths above 950 nm, the responsivity of the CCD falls rapidly, which limits the long wavelength NIR system response cutoff to 975 nm. Similarly, the short wavelength system response cutoff for the blue filter is 400 nm because of fall-off of the CCD responsivity at shorter wavelengths.**

*MVIC spectral response:* The spectral resolution of the MVIC channels is much lower than for LEISA and the CCD response varies relatively slowly, so a coarser measurement of the wavelength dependence of the filter/CCD assembly is acceptable. Figure 9 shows the measured sensitivity of the combined array/filter focal plane assembly. These measurements were obtained at 50 nm intervals using a broad spectral source filtered with 50 nm wide spectral filters as an input. In order to account for observational sources with greater spectral variation, the spectral responses of the filters were determined at higher spectral resolution and finer point spacing. Figure 10 shows the spectral response curves for the MVIC color filters. In addition to allowing for the retrieval of more accurate spectral models, these data will be used to calculate the sensitivity of MVIC to the source spectral distribution (e.g. a solar reflectance spectrum vs. the blackbody spectrum of a volcano on Io).

**5.2 Full Instrument Level Measurements**

*Absolute Radiometry*: A spectrally calibrated integrating sphere filling about a 1° field of view was used as the radiometric source. Radiometric response over the full MVIC image plane was measured by rotating the sphere illumination using a cryomirror assembly in the chamber. This technique was used to determine accurate relative radiometry (flatfields). The MVIC absolute radiometry was determined with an accuracy of approximately ±30%. The accuracy of the LEISA absolute radiometry was lower because water contamination in the integrating sphere caused significant absorption at some wavelengths. Radiometric calibration is being performed in-flight and the combined results of the pre-flight and in-flight calibration are shown in the next section.

*Image Quality*: The image quality, defined in terms of system MTF (Modulation Tranfer Function), was determined using a collimated point source to simulate a distant object. Collimation was verified interferometrically, as were corrections for pressure and thermally induced optical power in the chamber window. By defocussing the point source in a controlled fashion, this system could also be used to determine the best focus position for the focal planes. At the best focus position, a point source produced a spot in the focal plane whose FWHM was 1.2 ±0.1 pixel. After the tests were completed, it was found that there was some optical power in the cryomirror that was not accounted for in determining the focus. However it was determined that for the Ralph system with its 650-mm focal length, the focus error caused by the cryomirror curvature was negligible. The pre-flight image quality measurements were used to determine an expected budget for the full system MTF that is shown in the next section.

*SIA Pointing*: The pointing direction and spatial distribution of focal plane illumination for the SIA were also determined in the instrument level tests. When illuminated by a source simulating the angular size of the sun at Pluto, the SIA produced a stable pattern in both the MVIC and LEISA focal planes that was insensitive to the exact source position over a range of 0.5° in each dimension.

## 6.0 COMBINED PRE-LAUNCH AND IN-FLIGHT INSTRUMENT CALIBRATION RESULTS

Calibration of Ralph is being carried out in flight. To date, standard stars and Jupiter observations are being used to determine image quality, to measure over-all system radiometric sensitivity and, for LEISA, to verify the spectral calibration. Additionally, dense star-fields are being used to determine overall optical distortion. A Jupiter gravity assist will occur in 2007 with closest approach on February 28. It will provide additional opportunities for measuring the flat field response, and will permit radiometric and spectral calibration. Afterwards, every year during the flight, there will be a 50-day checkout period during which calibration will be checked. Extensive calibration will also be carried out during the 6-month period prior to and after the planned Pluto encounter on 14 July, 2015. Because the initial in-flight calibration analyses are not complete, the following results should be considered to be preliminary estimates.

So far in flight, the MVIC and LEISA system random noises have remained at their pre-launch values: ≤ 1 count (≤60 e-) for all the MVIC CCDs and 4.5 counts (50 e-) for LEISA. In addition to the random noise, the MVIC CCDs have some periodic noise (as much as 2 counts) that may be removed by post-processing. This noise was present before launch. The MVIC dark current is completely negligible; it is not measurable for the integration times allowed by the Ralph electronics. The LEISA dark current is ~40 counts/second (~440 e-/sec), and does not contribute significantly to the noise even for the longest allowed integration time (4 sec.). So far, the Ralph decontamination heaters have been left on except for 24-hour periods around data collection events in order to minimize the condensation of contaminants produced by spacecraft outgassing and thruster operation. This means that, as of this writing, the LEISA focal plane has not reached its quiescent operating temperature when acquiring data. A cool-down period of greater than 24 hours is common in low temperature, passively cooled systems. When the heaters are left off for longer periods, it is expected that the focal plane temperature will drop a few more degrees. The dark current will decrease by about a factor of two for every five degrees drop in the focal plane temperature.

### 6.1 Image Quality and MTF

The point spread function of MVIC's panchromatic TDI channels determined in-flight is well represented by a 2-D gaussian function. Fittings of the point source intensity distribution to this PSF for about 30 relatively bright stars has yielded FWHMs of 1.48±0.13 pixels in the in-track direction and 1.40±0.1 pixels in the cross-track direction. These observations are all for integration times of 0.5 seconds or less, for which the contribution of the uncorrected spacecraft motion is less than half a pixel. Taking the Fourier transform of the fitted PSF yields the expression:

$$\text{MTF}(d) = e^{-(d\pi 0.01977 \text{FWHM}/(2(\ln 2)^{1/2}))^2} \tag{1}$$

for the value of the MTF at spatial frequency, d (cycles/milliradian). Using this expression, MTF(20) = 0.30±0.07 (in-track) and MTF(20) = 0.34±0.06 (cross-track). Both the in-track and cross-track MTF(20) values surpass the requirement that they must be greater than 0.15 (see Table 1). These results are summarized in Figure 11, which shows the pre-launch MTF curves, determined solely by the optical characteristics, and the pre-flight model of the full system. In this figure the top area in each panel ("measured spots") represents the measured contribution of the instrumental characteristics to the MTF. Off array integration is a term that accounts for the effect of insufficient masking of the bulk silicon from the light. The other factors are based on predicted spacecraft behavior at the 3σ level. The temporal aperture and the TDI error have to do with uncompensated spacecraft motion. The point labeled "requirement" is the 15% minimum overall MTF at 20 cycles/milliraian (0.4 cycles/pixel) that satisfies the science requirements. The measured in-flight image characteristics clearly exceed requirements and agree remarkably well with the predicted behavior based on the pre-flight instrument level measurements.

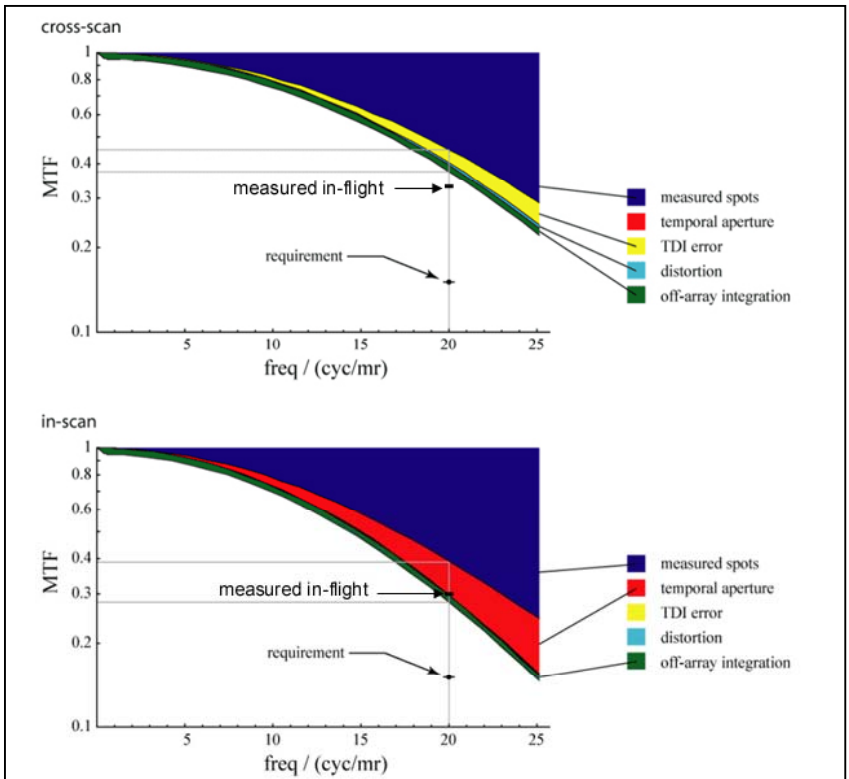

Figure 11: Modeled and measured MTF for the MVIC panchromatic TDI channel. "Measured spots" represents the measured contribution of the optical characteristics to the MTF, while the other factors are based on predicted behavior at the 3σ level. The temporal aperture and the TDI error have to do with uncompensated spacecraft motion. Off array integration is a term that comes about because of insufficient masking of the bulk silicon from the light. The points labeled "measured in-flight" are derived from stellar observations and represent the full system MTF. These clearly exceed the requirement.

The MVIC color channels and the LEISA spectral channels have no additional requirements on image quality beyond the requirement that they be in focus to within 0.5 pixel. This is because the required spatial resolutions for the observations using the MVIC color and LEISA channels are significantly lower than for the MVIC panchromatic channel. The MVIC color channels are at the same focal distance as the panchromatic channel by virtue of being on the same substrate. The LEISA focal plane was focused separately. Table 3 summarizes the preliminary results of the analysis of the PSF for all the Ralph channels. The effect of diffraction is apparent in the FWHM measurements for the MVIC color channels. The LEISA results are based on a single stellar observation that only covered from 1.25 to 1.8 μm. Diffraction effects are not apparent in these LEISA measurements because the LEISA pixels are three times larger than the MVIC pixels.

## 6.2 Optical Distortion

The MVIC focal plane covers 5.7° in the cross track direction, which means that it is expected that there will be optical distortion of on the order of a few pixels at the ends of the FOV. In addition, the three-mirror anastigmat design is known to be anamorphic. That is, the spatial scale in the cross track direction is slightly different from that in the along track direction. These effects both make the apparent position of objects different from their true position. In order to account for this effect, both to maintain spatial fidelity in the science data and to allow accurate observations for optical navigation, a distortion correction must be found that maps the apparent position to the true position. The distortion effect is illustrated in Figure 12 for a set of observations of a dense star field using MVIC's panchromatic framing camera. This plot shows the difference between the true position of a star and the position obtained assuming that each pixel is 19.77 μm square. The stellar positions are typically known to better than 5 μrad. As may be seen from this figure, there is about a three-pixel distortion at the ends of the focal plane. Note that the column distortion is sinusoidal about the center of the FOV, while the row distortion is the same sign on opposite sides of the array. The anamorphism is the principle driver for the row distortion. These results are preliminary and are being refined further, but they indicate the effects are easily modeled.

**Table 3: Summary of Full Width at Half Maximum for Ralph Channels[1]**

| Channel | In-Track FWHM | Cross-Track FWHM |
|---|---|---|
| MVIC Pan | 1.48±0.13 | 1.40±0.10 |
| MVIC Blue | 1.48±0.10 | 1.29±0.07 |
| MVIC Red | 1.55±0.12 | 1.38±0.08 |
| MVIC NIR | 1.95±0.15 | 1.97±0.15 |
| MVIC $CH_4$ | 1.78±0.13 | 1.81±0.20 |
| [2]LEISA | 1.40±0.13 | 1.40±0.10 |

[1]Initial results for FWHM in units of pixels.
[2]This is the result of a single stellar observation and is the average FWHM for wavelengths from 1.25 to 1.8 μm. There is no apparent wavelength dependence over this range.

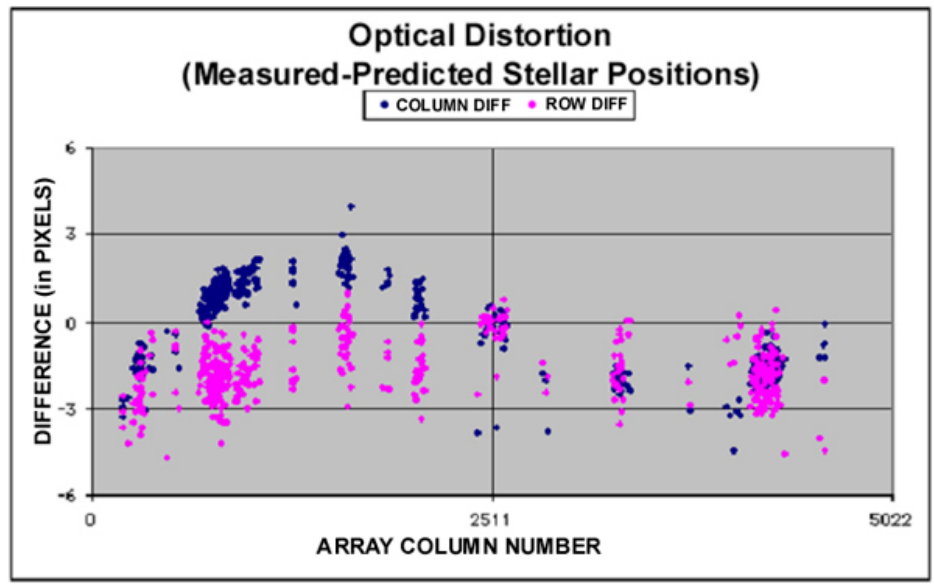

**Figure 12: Effect of optical distortion for MVIC framing camera. The plotted points are the differences between predicted row and column stellar positions and true row and column stellar positions as determined using dense star fields. The differences are plotted as a function of array column number.**

### 6.3 Radiometric Calibration

The absolute radiometric calibration of the MVIC and LEISA components of Ralph is primarily being carried out in-flight, using point sources (standard stars) and Jupiter. The flatfields determined during the pre-launch testing are used to interpolate the stellar and Jupiter results over the entire focal planes. This work is still progressing. The results below are preliminary, and are expected to be accurate to about 10%. Figure 13 shows an example of the stellar source calibration measurements for the MVIC framing array. This figure shows a plot of the integrated signal at the focal plane *vs* the visual magnitude of a star, for a number of A-, B- and K-type stars. A fit was done using data for B9-type stars that is linear to a very high degree of accuracy (it accounts for 99.5% of the standard deviation of the data). The other A- and B- type stars are well described by this fit as well. The colder, K-type star is a bit of an outlier, but no correction has been made for color temperatures of the stars. These data indicate that, for the framing array, a 14th magnitude B9 star will produce an integrated signal of one count per second. Given the MVIC frame noise characteristics (~ 0.6 counts of noise at low illumination levels), a tenth magnitude star will produce a measurement with a signal-to-noise ratio of 14 in 0.25 seconds if it is focused onto a single pixel. If the same signal is spread out over 4 adjacent pixels it produces an average SNR of 8, which still meets the OpNav requirement.

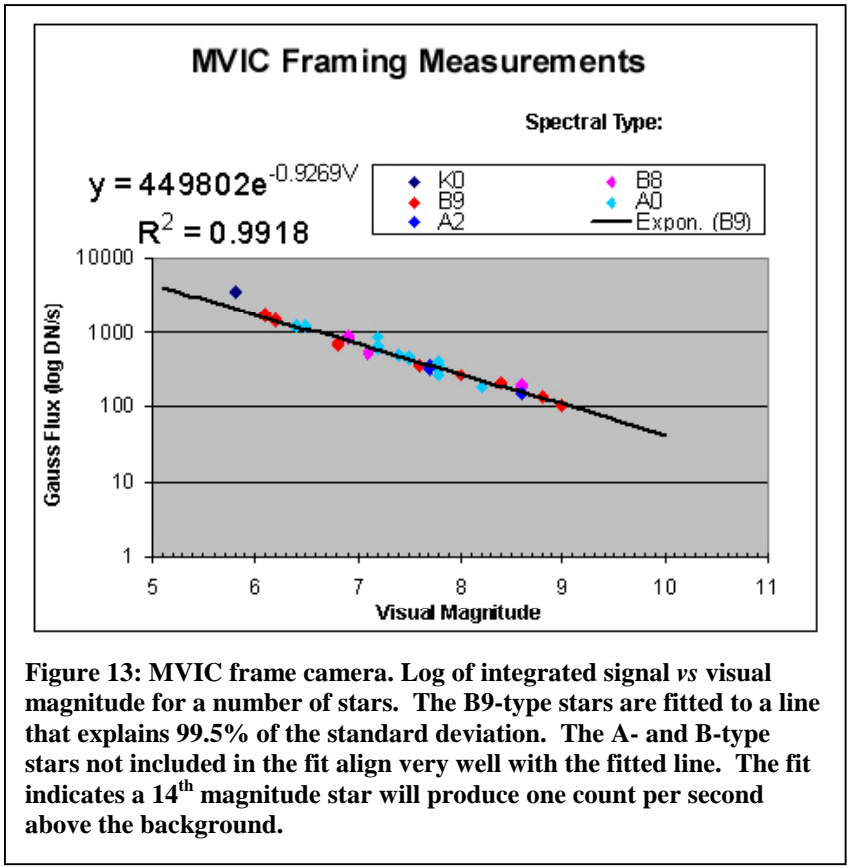

**Figure 13: MVIC frame camera. Log of integrated signal *vs* visual magnitude for a number of stars. The B9-type stars are fitted to a line that explains 99.5% of the standard deviation. The A- and B-type stars not included in the fit align very well with the fitted line. The fit indicates a 14th magnitude star will produce one count per second above the background.**

Table 4 shows the sensitivities and the predicted signal-to-noise ratios for the MVIC channels and for the LEISA wavelengths at which the LEISA SNR performance requirements were defined. The predictions are made for the conditions specified in Table 1 which are representative of the flux levels expected at Pluto. The table also lists the noise performance requirements set forth in the Announcement of Opportunity (AO) for the mission. As can be seen from this table, all MVIC channels easily meet their sensitivity requirements. Similarly, LEISA meets the AO radiometric performance specifications, but with significantly smaller margin, particularly at shorter wavelengths. The decreased performance in this spectral region is caused by a known drop in quantum efficiency of the array at wavelengths shorter than 1.6 μm, and by lower transmittance of the filter at shorter wavelengths. However, solar flux increases at the shorter wavelengths so the decreased efficiency is compensated for by

| Table 4: Ralph Radiometric Sensitivity, Predicted Signal-to-Noise Ratios and Mission Requirements | | | |
|---|---|---|---|
| **Channel** | **Sensitivity (DN/photon)[1]** | **Predicted SNR[2]** | **Required SNR** |
| MVIC Pan | $1.89 \times 10^{-3}$ | 150 | 50 |
| MVIC Blue | $1.38 \times 10^{-3}$ | 68 | 50 |
| MVIC Red | $2.25 \times 10^{-3}$ | 122 | 50 |
| MVIC NIR | $1.59 \times 10^{-3}$ | 106 | 50 |
| MVIC $CH_4$ | $2.00 \times 10^{-3}$ | 48 | 15[3] |
| [4]LEISA 1.25 μm | $2.45 \times 10^{-3}$ | 32 | 31 |
| [4]LEISA 2.00 μm | $5.82 \times 10^{-3}$ | 35 | 27 |
| [4]LEISA 2.15 μm | $8.32 \times 10^{-3}$ | 24 | 18 |

[1] For MVIC, there are on the average, 58.6 e- per DN, where DN (digital number) is the least significant bit of the A/D. For LEISA there are about 11 e- per DN.
[2] Predicted SNR at Pluto. MVIC: Assumes 35% albedo; LEISA: Assumes Pluto model albedo in Figure 1. MVIC Pan is for 0.4 sec integration, MVIC color is for 0.6 sec integration, LEISA is for 0.5 sec integration.
[3] The methane specification is an internal goal, not a requirement
[4] The LEISA SNRs are for the average of two planned scans

the larger flux. Figure 14 shows the LEISA radiometric sensitivity as a function of wavelength. Figure 15 shows the predicted signal-to-noise ratio of LEISA observations (average of two scans, 0.5 second /pixel integration time) of Pluto and Charon for the nominal albedos plotted in Figure 1. At the Pluto flux levels the noise is dominated by the system noise (read noise), so that the signal-to-noise ratio grows nearly linearly with increased integration time.

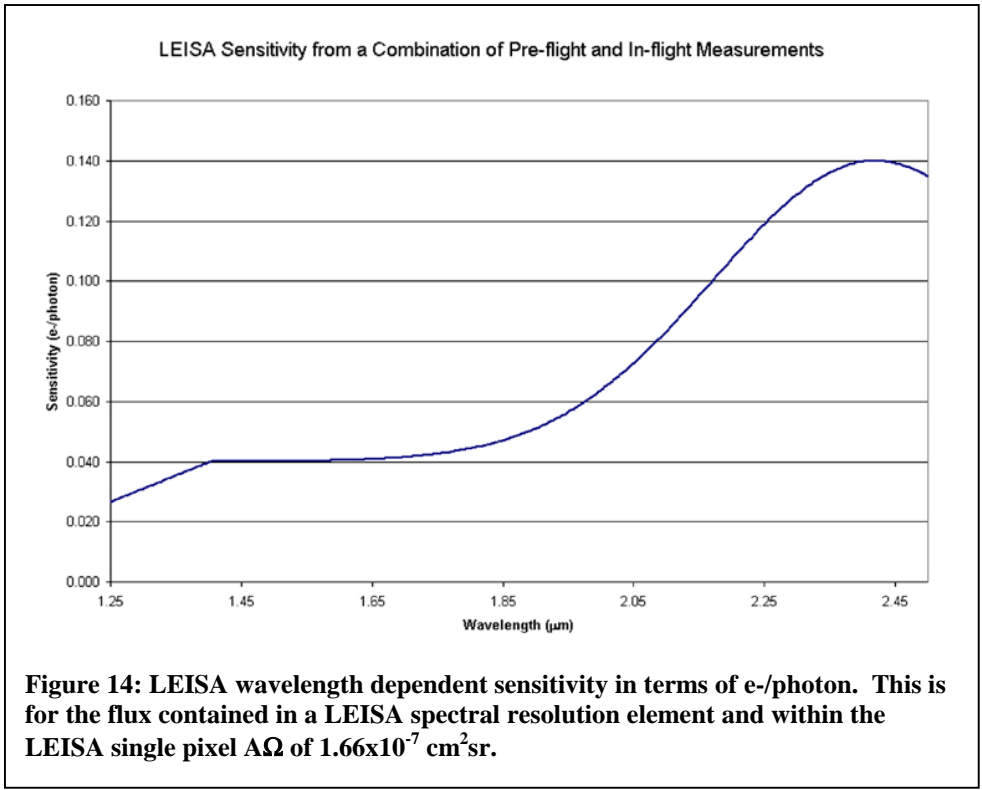

**Figure 14: LEISA wavelength dependent sensitivity in terms of e-/photon. This is for the flux contained in a LEISA spectral resolution element and within the LEISA single pixel AΩ of $1.66 \times 10^{-7}$ cm$^2$sr.**

### 6.4 Anomalous Solar Light Leak

In-flight testing of Ralph has shown the presence of an anomalous background signal in the LEISA imager that appears to be caused by the transmission of a very small fraction of the ambient solar flux into the area behind the focal plane. These photons pass between the filter and the array and give rise to a "solar light leak" signal that, in the worst case, is less than 1 part in $10^7$ of the ambient solar flux. The background may be eliminated by using structures on the spacecraft to shield Ralph from the sun. This behavior, and the $R_{sun}^2$ dependence of the magnitude of the background signal are both evidence of the sun as the source of this anomaly. Ralph can not always be shielded. When Ralph is not shielded, the magnitude of the effect is a slowly varying function of the position of the sun relative to Ralph and it may be removed to a high degree of accuracy. This means that the primary result of the light leak is to increase the system noise because of photon counting statistics. At Pluto's heliocentric distance, and for the integration times that are possible with LEISA, the excess background introduced by the effect is at least a factor of two less than the read-noise equivalent input flux. Therefore, after the background is removed, there will be little or no measurable effect of this excess flux on the LEISA observational results. As of yet, the root manufacturing cause of the light leak has not been determined. It is possible that light is more of less uniformly penetrating the multi-layer insulation (MLI) thermal shield that encloses the instrument, and propagating along LEISA's interface cables to the focal plane. However, definitive answers, if they can be determined, await further laboratory and in-flight measurements. Nevertheless, for practical purposes, the phenomenology is well understood. MVIC shows no evidence of this effect.

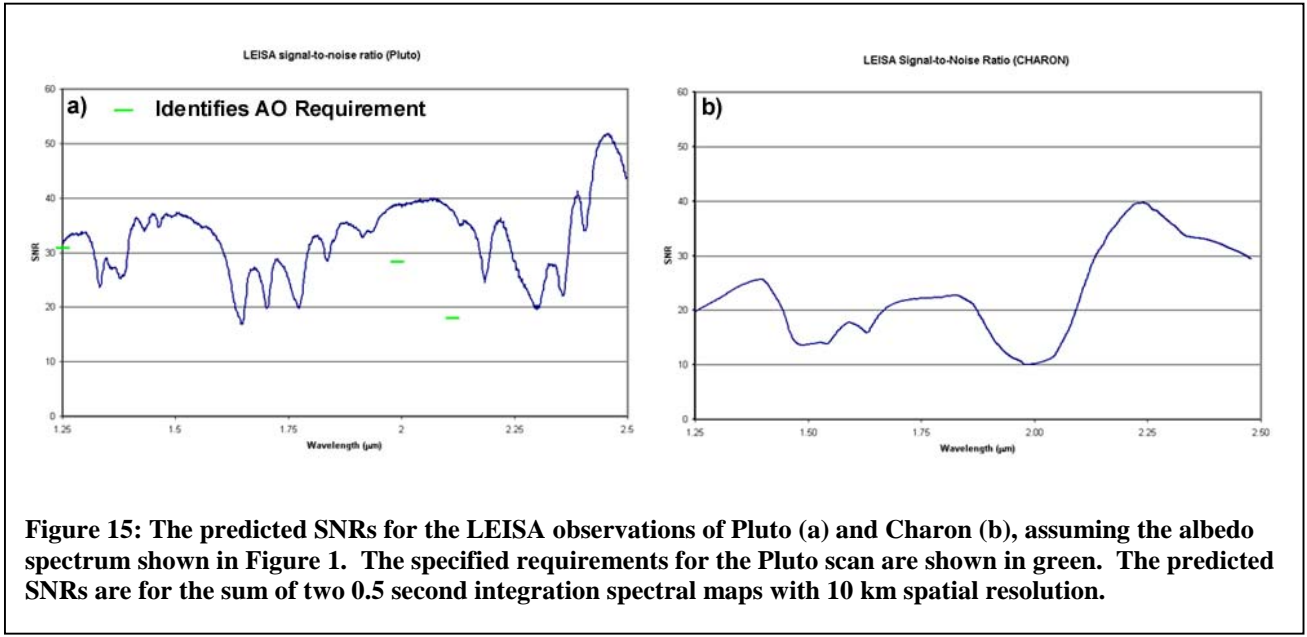

**Figure 15: The predicted SNRs for the LEISA observations of Pluto (a) and Charon (b), assuming the albedo spectrum shown in Figure 1. The specified requirements for the Pluto scan are shown in green. The predicted SNRs are for the sum of two 0.5 second integration spectral maps with 10 km spatial resolution.**

## 7.0 IN-FLIGHT INSTRUMENT OPERATION

Ralph data collection operations fall into one of four categories: 1) panchromatic MVIC TDI, 2) color MVIC TDI, 3) LEISA and 4) panchromatic MVIC framing. Ralph may operate in only one of these categories at any time meaning, for example, that LEISA data may not be taken simultaneously with MVIC data. The panchromatic MVIC TDI category covers operation in either of the panchromatic TDI CCD arrays, but only one pan array can be operated at any time. All four color TDI CCDs operate simultaneously. As a result, a color MVIC scan is slightly longer than the equivalent Pan TDI scan, because the target must be scanned over all 4 color CCDs. The category is chosen by command and implemented by a set of relays.

For the MVIC TDI and LEISA data categories operation may be either in calculated rate mode or forced rate mode. In calculated rate mode, the array read-out rate is set using the measured rotation rate about the scan axis (the Z-axis) as determined by the spacecraft's star trackers and gyroscopes. This scan rate is provided to Ralph at the beginning of a scan by the spacecraft's guidance and control (G&C) system. The array readout rate is set such that the scan moves a spot on the image a single row between reads. In calculated rate mode, the MVIC TDI arrays may scan only in one direction, but LEISA may scan in either direction. In forced rate mode, the array readout rate is set by a command and is not coupled to the G&C scan rate. For all categories, data collection is initiated with a start command and continues until a stop command is received. MVIC framing data is always obtained in forced rate mode.

There are two types of scans. The normal scan type is used when the target is sufficiently distant that the effective scan rate of the boresight caused by the relative motion of the target and the spacecraft remains constant during the scan. For a normal scan, once the rotation rate has stabilized, the along track thrusters are disabled and the spacecraft is allowed to rotate at a constant rate. In calculated rate mode the rotation rate, measured to a $3\sigma$ accuracy of $\pm 7\mu rad/sec$ is passed to Ralph and the calculated frame rate is matched to the boresight motion. A normal type scan using calculated rate mode is the most common form of either MVIC TDI or LEISA operation. Typically, in a normal LEISA scan, the cross track thrusters are enabled so that the target does not drift off the focal plane in the direction perpendicular to the scan during the data collection event. For a given target, the MVIC scans are usually almost an order of magnitude faster than the LEISA scans, so the cross track thrusters are typically disabled in MVIC TDI scans. Correct operation of the normal scan type has been verified in flight for both MVIC and LEISA using stellar sources.

When the target is sufficiently close that the effective rotation rate induced by the relative motion of the target and the spacecraft changes during the scan, the scan becomes more complex. In this case, the rotation rate of the spacecraft is changed during the scan by thruster firings. This type of scan, called a pseudo CB3 scan, is implemented by making the boresight track an artificial object whose ephemeris is defined in such a fashion to keep the combined boresight rotation rate constant. The frame rate is set by the commanded scan rate, and not by the changing spacecraft rotation rate as measured by the G&C system. The scan is controlled to within $\pm 34\mu rad/sec$. Pseudo CB3 scans are only used when the target is close to the spacecraft, such as the near closest approach LEISA and MVIC Pluto scans. For this type of operation, both the in track and cross track thrusters are enabled during LEISA scans, while only the in track thrusters are enabled during MVIC scans. Correct operation of the pseudo CB3 scan type has been verified in flight for both MVIC and LEISA using an asteroid near encounter (Olkin *et al*, 2007).

## 8.0 CONCLUSION

This paper describes the design, operation and performance of Ralph, a highly capable remote sensing imager/IR spectral imager flying on the New Horizons mission to the Pluto/Charon system and the Kuiper belt beyond. Ralph consists of a telescope feeding two focal planes, the visible/NIR MVIC imager and the LEISA IR spectral imager. MVIC will provide very sensitive, high fidelity, full hemispheric panchromatic maps of Pluto and Charon at a spatial resolution of 1 km/linepair and VIS/NIR color maps at a spatial resolution of better than 4km/linepair. LEISA will provide full hemispheric, SWIR, spectral maps with spatial resolutions of 7 km/pixel or better. These will be used to accurately characterize the surface composition. At closest approach, MVIC will obtain images with spatial resolutions on the order of a few hundred meters in selected areas, while LEISA will measure spectra at the 1-2 km spatial scale. Ralph has been extensively tested at the component and full instrument level and its in-flight operation has verified that it meets all its performance requirements with margin. Ralph will provide a wealth of information on the geology, composition, morphology and thermal characteristics of the Pluto/Charon system. The data it produces during its flyby will revolutionize our understanding of Pluto and its neighbors and will shed new light on the evolution of our solar system and the nature of the objects in the Pluto system.

Eight years prior to arriving at Pluto, Ralph will observe Jupiter for a period of about 4 months during its close approach for a gravity assist, starting in January of 2007. The closest approach of ~33 $R_J$ will occur on 28 February, 2007. During this period, numerous Ralph observations of Jupiter and its moons are planned to provide encounter practice and to obtain calibration measurements on the last fully resolved object prior to the Pluto encounter 8 years hence. The LEISA observation obtained at this time will be some of the highest spectral/spatial data ever obtained of Jupiter in the SWIR spectral range. Thus New Horizons will provide exciting new science, even before its rendezvous with Pluto.

## ACKNOWLEDGEMENTS

The authors would like to thank the entire Ralph support teams at BATC and SwRI and the LEISA support team at GSFC for their untiring efforts in making Ralph a reality. The support of JDSU/Uniphase, E2V, Barr Associates and Rockwell Scientific Corporation is also gratefully acknowledged.

## 9.0 REFERENCES

Cruikshank, Dale and Cristina M. Dalle Ore (private communication, 2000).

Grundy, W. M., B. Schmitt and E Quirico, "The Temperature Dependent Spectra of Alpha and Beta Nitrogen Ice with Application to Triton." Icarus, 105, 254, 1993.

Olkin, C. B., D. C. Reuter, A. Lunsford, R. P. Binzel, S. A. Stern, "The New Horizons Distant Flyby of Asteroid 2002 JF56.", in preparation

Reuter, D. C., D. E. Jennings, G. H. McCabe, J. W. Travis, V. T. Bly, A. T. La, T. L. Nguyen, M. D. Jhabvala, P. K. Shu and R. D. Endres, "Hyperspectral Sensing Using the Linear Etalon Imaging Spectral Array." *SPIE Proceedings of the European Symposium on Satellite Remote Sensing III: Conference on Sensors, Systems, and Next Generation Satellites II,* **2957**, 154-161, September 23-26, 1996, Taorima, Sicily, Italy.

Reuter, D. C., G. H. McCabe, R. Dimitrov, S. M. Graham, D. E. Jennings, M. M. Matsumura, D. A. Rapchun and J. W. Travis, "The LEISA/ Atmospheric Corrector (LAC) on EO-1", *IGARS Proceedings; IEEE 2001 International Geoscience and Remote Sensing Symposium*, Volume 1, 46–48, July 9–13, 2001, Sydney, Australia.

Reuter, Dennis, Alan Stern, James Baer, Lisa Hardaway, Donald Jennings, Stuart McMuldroch, Jeffrey Moore, Cathy Olkin, Robert Parizek, Derek Sabatke, John Scherrer, John Stone, Jeffrey Van Cleve and Leslie Young, paper 5906-51, "Ralph: A visible/infrared imager for the New Horizons Pluto/Kuiper Belt Mission", *SPIE Proceedings of the Optics and Photonics Conference, Astrobiology and Planetary Missions*, Vol. 5906, 59061F-1 to 59061F-11, July 31- August 4, 2005, San Diego CA.

Rosenberg, K. P., K. D. Hendrix, D. E. Jennings, D. C. Reuter, M. D. Jhabvala, and A. T. La, "Logarithmically Variable Infrared Etalon Filters.", *SPIE Proceedings, Optical Thin Films IV: New Developments*, **2262**, 25 - 27 July, 1994, San Diego, CA.

Stern, S. A., D. C. Slater, W. Gibson, H. J. Reitsema, Alan Delamere, D. E. Jennings, D. C. Reuter, J. T. Clarke, C. C. Porco, E. M. Shoemaker and J. R. Spencer, "The Highly Integrated Pluto Payload System (HIPPS): A Sciencecraft Instrument for the Pluto Mission.", *SPIE Proceedings, EUV, X-RAY and Gamma-Ray Instrumentation for Astronomy VI*, **2518**, 39 - 58, San Diego, CA; July 1995.

Ungar, S. G., J. S. Pearlman, J. A. Mendenhall, D. Reuter, "Overview of the Earth Observing One (EO-1) mission", *IEEE Transactions on Geoscience and Remote Sensing*, **41** (part 1), 1149-1159, 2003.